\documentstyle[12pt,a4wide,draft,bbold]{article}

\def\bea{\begin{eqnarray}}
\def\eea{\end{eqnarray}}
\def\be{\begin{equation}}
\def\ee{\end{equation}}
\def\nn{\nonumber}
\def\ds{\partial\!\!\! /}
\def\dbs{\bar\partial\!\!\! /}
\def\uni{{\mathbb{1}}}

\def\tr{\mbox{tr}}
\newcommand{\NPB}[3]{{Nucl.\ Phys.} {\bf B#1} (#2) #3}

\newcommand{\JHEP}[3]{{JHEP} {\bf #1} (#2) #3}
\newcommand{\hepth}[1]{{\tt hep-th/#1}}
\newcommand{\ft}[2]{{\textstyle\frac{#1}{#2}}\,}
\newcommand{\eqn}[1]{(\ref{#1})}

\begin{document}

\thispagestyle{empty}
\begin{flushright}
{\small hep-th/0002107}\\
{\small AEI-2000-004}\\[3mm]
\end{flushright}

\vspace{1cm}
\setcounter{footnote}{0}
\begin{center}
{\Large{\bf The Asymptotic Groundstate of SU(3) Matrix Theory }
    }\\[14mm]

 {\sc  Jens Hoppe and Jan Plefka}\\[10mm]

{\em Albert-Einstein-Institut}\\
{\em Max-Planck-Institut f\"ur
Gravitationsphysik}\\
{\em Am M\"uhlenberg 1, D-14476 Golm, Germany}\\
{\footnotesize \tt hoppe,plefka@aei-potsdam.mpg.de}\\[7mm]

{\sc Abstract}\\
\end{center}

The asymptotic form of a $SU(3)$ matrix theory groundstate 
is found by showing that a recent ansatz for a 
supersymmetric wavefunction is non-trivial (i.e. non-zero).

\vfill
\leftline{{\sc February 2000}}

\newpage
\setcounter{page}{1}


Maximally supersymmetric $SU(N)$ gauge quantum mechanics \cite{CH} in $d=9$
has in recent years received much attention due to its close 
relation\footnote{based on \cite{jensMIT}}
to the eleven-dimensional supermembrane \cite{DWHN}
in the $N\rightarrow\infty$ limit, its description
of the dynamics of $N$ D0 branes in superstring theory \cite{d0}, 
as well as the M theory proposal of \cite{BFSS}. In these physical
interpretations the existence of a unique normalizable zero-energy 
groundstate \cite{groundstate} is an important consistency 
requirement. An explicit construction of the
vacuum state, though highly desirable, appears to be quite difficult.
Another approach is to study the behavior of the wavefunction
far out at infinity where the degrees of freedom in the
Cartan-subalgebra become free and the remaining 
degrees of freedom form the zero energy vacuum state of  
supersymmetric harmonic oscillators \cite{ags,su2}. The full
asymptotic groundstate was constructed for the $SU(2)$ model in 
\cite{su2}, here we consider the $SU(3)$ case. 
Assuming that the Cartan-subalgebra degrees of
freedom are asymptotically governed by a set of free effective
supercharges $Q_\alpha$ a proposal was recently made \cite{jens}
as to which of the harmonic wavefunctions constructed in \cite{bhs}
is annihilated by the $Q_\alpha$. In this letter we prove the
non-triviality of this ansatz.


The asymptotic supersymmetry charge for the $d=9$ SU(3) model reads
\be
Q_\alpha=-i \Gamma_{\alpha\beta}^a\, \Bigl( \theta^1_\beta\, 
\frac{\partial}{\partial x_1^a} + \theta^2_\beta\, 
\frac{\partial}{\partial x_2^a} \Bigr)
\label{Q}
\ee
where $x_1^a,x_2^a$ $(a=1,\ldots,9)$ are the bosonic and
$\theta^1_\alpha,\theta^2_\alpha$ $(\alpha=1,\ldots,16)$ are
the fermionic degrees of freedom of the Cartan sector; we work
with a real, symmetric representation of the Dirac matrices 
and our charge conjugation matrix equals unity. It is advantageous
to go to the complex variables
\bea
\lambda= \ft{1}{\sqrt 2} (\theta^1 + i\theta^2) &\qquad&
z^a= \ft{1}{\sqrt 2} (x^a_1 + ix^a_2) \nn\\
\lambda^\dagger= \ft{1}{\sqrt 2} (\theta^1 - i\theta^2)  &\qquad&
\bar z^a= \ft{1}{\sqrt 2} ( x^a_1 - ix^a_2) \, .
\eea
Note that we have now divided the fermions into creation and
annihilation operators, obeying the algebra
\be
\{\lambda_\alpha, \lambda^\dagger_\beta\}= \delta_{\alpha \beta}\, ,
\ee
and we define the fermionic vacuum $|-\rangle$  by
$\lambda_\alpha\, |-\rangle =0$. The completely filled state is 
denoted by $|+\rangle=\ft{1}{16!}\epsilon^{\alpha_1\ldots\alpha_{16}}
\, \lambda^\dagger_{\alpha_1}\ldots\lambda^\dagger_{\alpha_{16}}|-\rangle$.
Clearly $|-\rangle$ and $|+\rangle$ are $SO(9)$ singlets. However, there
is a third $SO(9)$ singlet state
\be
|\uni\rangle = 
(\lambda^\dagger \Gamma^{ab}\lambda^\dagger)\,
(\lambda^\dagger \Gamma^{bc}\lambda^\dagger)\,
(\lambda^\dagger \Gamma^{cd}\lambda^\dagger)\,
(\lambda^\dagger \Gamma^{da}\lambda^\dagger)\, |-\rangle
\label{uni}
\ee
in the half-filled sector. It can be shown \cite{bhs} 
that there are no further $SO(9)$ singlets. A further symmetry
group acting on these states is the Weyl group, the discrete 
asymptotic remnant of the continous $SU(3)$ of the full system.
The Weyl group for $SU(3)$ may be generated by two elements 
$P$ and $C$ \cite{bhs}, which act on the complex fermions
$\lambda$ and $\lambda^\dagger$ as
\bea
P: \qquad \lambda \rightarrow & \lambda^\dagger 
&\qquad \lambda^\dagger \rightarrow \lambda
\nn\\
C: \qquad \lambda \rightarrow& e^{-\ft{2\pi i}{3}}\, \lambda 
&\qquad \lambda^\dagger \rightarrow e^{\ft{2\pi i}{3}}\lambda^\dagger \,.
\eea
As $P$ interchanges $|+\rangle$ and $|-\rangle$, leaves the eight
fermion sector invariant and the three $SO(9)$ singlets are known to
form one two dimensional irreducible representation under the Weyl group
and one singlet \cite{jens} the state $|\uni\rangle$ has to be 
Weyl invariant. This is consistent with $C$ transforming
$|\pm\rangle$ into $\exp(\mp\ft{2\pi i}{3})\, |\pm\rangle$.
So $|\uni\rangle$ is the unique $SO(9)$ {\it and} Weyl invariant state.

In the complex variables the supersymmetry charge \eqn{Q} reads
$Q_\alpha = -i(\ds \lambda)_\alpha -i(\dbs \lambda^\dagger)_\alpha$,
where $\partial_a= d/d\, z^a$ and $\bar\partial_a= d/d\, {\bar z}^a$.
We seek for an asymptotic groundstate $|\Psi\rangle$ obeying
$Q_\alpha|\Psi\rangle =0$. Note that while $Q_\alpha$ 
squares to $-\partial\cdot\bar\partial$, the condition
$\partial\cdot\bar \partial\, |\Psi\rangle=0$ does not imply
$Q_\alpha\, |\Psi\rangle=0$ due to the purely asymptotic considerations,
i.e. $|\Psi\rangle$ not being square integrable caused by its singularity
at the origin.

Consider now the ansatz for $|\Psi\rangle$
\be
|\Psi\rangle = \epsilon^{\alpha_1\ldots \alpha_{16}}\,
Q_{\alpha_1}\ldots Q_{\alpha_{16}}\, \frac{1}{(z\cdot \bar z)^8}\, |\uni\rangle
\, .
\label{ansatz}
\ee
$|\Psi\rangle$ is obviously annihilated by $Q_\alpha$, as $Q_\alpha$
squares to the Laplacian $\partial\cdot\bar\partial$ which in turn
annihilates the harmonic function $(z\cdot \bar z)^{-8}$. 
Note that $|\Psi\rangle$ is $SO(9)\times$Weyl invariant by construction.
What remains to be shown, however, is that $|\Psi\rangle$ is 
{\it non-vanishing}.

For this we consider the matrix element
\bea
\langle -| \, \Psi\rangle &=&
\epsilon^{\alpha_1\ldots \alpha_{16}}\,\langle -| \, 
[(\ds \lambda)_{\alpha_1} +(\dbs \lambda^\dagger)_{\alpha_1}]\ldots 
[(\ds \lambda)_{\alpha_{16}} +(\dbs \lambda^\dagger)_{\alpha_{16}}]
\nn\\ && \qquad \qquad \qquad 
(\lambda^\dagger \Gamma^{ab}\lambda^\dagger)\,
(\lambda^\dagger \Gamma^{bc}\lambda^\dagger)\,
(\lambda^\dagger \Gamma^{cd}\lambda^\dagger)\,
(\lambda^\dagger \Gamma^{da}\lambda^\dagger)\, |-\rangle
\, \frac{1}{(z\cdot \bar z)^8}\, ,
\label{me}
\eea
which we now need to normal order by making use of the anticommutator
relation
\be
\{(\ds \lambda)_\alpha, (\dbs \lambda^\dagger)_\beta \} =
\delta_{\alpha\beta}\, \partial\cdot\bar\partial + \Gamma_{\alpha\beta}^{ab}\,
\partial_a\, \bar\partial_b \, .
\ee
From the $2^{16}$ terms generated from expanding out the brackets in
the first line of \eqn{me} only those containing 4 $(\dbs \lambda^\dagger)$
and 12 $(\ds \lambda)$ survive. Normal ordering of these terms then
yields
\bea
\langle -| \, \Psi\rangle &\sim&
\epsilon^{\alpha_1\ldots \alpha_{16}}\,
\Gamma^{a_1 a_2}_{\alpha_1\alpha_2}\,\bar\partial_{a_1}\partial_{a_2}\,
\Gamma^{a_3 a_4}_{\alpha_3\alpha_4}\,\bar\partial_{a_3}\partial_{a_4}\,
\Gamma^{a_5 a_6}_{\alpha_5\alpha_6}\,\bar\partial_{a_5}\partial_{a_6}\,
\Gamma^{a_7 a_8}_{\alpha_7\alpha_8}\,\bar\partial_{a_7}\partial_{a_8}\nn\\
&& 
\langle -| \, 
(\ds\lambda)_{\alpha_9}\ldots(\ds\lambda)_{\alpha_{16}}\,
(\lambda^\dagger \Gamma^{ab}\lambda^\dagger)\,
(\lambda^\dagger \Gamma^{bc}\lambda^\dagger)\,
(\lambda^\dagger \Gamma^{cd}\lambda^\dagger)\,
(\lambda^\dagger \Gamma^{da}\lambda^\dagger)\, |-\rangle
\, \frac{1}{(z\cdot\bar z)^8}\, , \nn
\label{me2}
\eea
and the final contractions then result in
\bea
\langle -| \, \Psi\rangle &\sim&
\epsilon^{\alpha_1\ldots \alpha_{16}}\,
\Gamma^{a_1 a_2}_{\alpha_1\alpha_2}\,\bar\partial_{a_1}\partial_{a_2}\,
\Gamma^{a_3 a_4}_{\alpha_3\alpha_4}\,\bar\partial_{a_3}\partial_{a_4}\,
\Gamma^{a_5 a_6}_{\alpha_5\alpha_6}\,\bar\partial_{a_5}\partial_{a_6}\,
\Gamma^{a_7 a_8}_{\alpha_7\alpha_8}\,\bar\partial_{a_7}\partial_{a_8}\nn \\
&&\qquad
(\ds \Gamma^{ab}\ds)_{\alpha_9\alpha_{10}}\,
(\ds \Gamma^{bc}\ds)_{\alpha_{11}\alpha_{12}}\,
(\ds \Gamma^{cd}\ds)_{\alpha_{13}\alpha_{14}}\,
(\ds \Gamma^{da}\ds)_{\alpha_{15}\alpha_{16}}
\, \frac{1}{(z\cdot \bar z)^8} \, ,
\label{me3}
\eea
where the precise (non-zero) combinatorial coefficient in this 
relation is not of 
interest, as we only need to show the non-vanishing of 
$\langle - |\, \Psi \rangle$. In order to proceed we note that
\be
(\ds \Gamma^{ab}\ds)_{[\alpha\beta]}= \Gamma_{\alpha\beta}^{ab}\, 
\partial\cdot\partial + 4\,\partial^{[a}\, {\Gamma^{b]c}}\!\!\!\!{}_{\alpha\beta}
\, \partial_c \, .
\label{gammaid}
\ee
Hence \eqn{me3} may be reduced to a differential operator in
$\partial_a$ and $\bar\partial_a$ of degree 16 acting on 
$(z\cdot \bar z)^{-8}$ 
provided we know the precise form of the 16 index tensor 
$t_{(16)}^{\,\, a_1\ldots a_{16}}$
\be
t_{(16)}^{\,\, a_1\ldots a_{16}}= \epsilon^{\alpha_1\ldots \alpha_{16}}\,
\Gamma_{\alpha_1\alpha_2}^{a_1a_2}\,
\Gamma_{\alpha_3\alpha_4}^{a_3a_4}\ldots
\Gamma_{\alpha_{15}\alpha_{16}}^{a_{15}a_{16}} \, .
\label{t16}
\ee
Clearly $t_{(16)}^{\,\, a_1\ldots a_{16}}$ 
must be expressable in form of a large string
of space-indexed $\delta$-functions, the $\epsilon^{a_1\ldots a_9}$ 
symbol cannot appear.
Interestingly enough this tensor also appears in the leading 
one-loop quantum correction to the M-theory effective action 
contracted with four Riemann tensors \cite{GKP}. 
Its precise form can be computed \cite{NP} and is most conveniently 
written down in a form contracted with an antisymmetric auxiliary 
tensor $X^{ab}$
\bea
t_{(16)}^{\,\, a_1\ldots a_{16}}\, X^{a_1 a_2}\ldots X^{a_{15}a_{16}}
&=& 105 \cdot 2^{19}\, \Bigl [ -5\, (\tr X^2)^4 + 384\, \tr X^8
- 256\, \tr X^2 \, \tr X^6
\nn\\&&\qquad\qquad
 + 72\, (\tr X^2)^2\, \tr X^4 - 48 \, (\tr X^4)^2\,
\Bigr ]
\label{t16res}
\eea
where the product of $X$ is to be understood in the matrix sense.

The knowledge of $t_{(16)}^{\,\, a_1\ldots a_{16}}$ now enables us to
finally evaluate \eqn{me3} using \eqn{gammaid}, which is still rather
involved and most effectively done with the help of the computer
algebra system FORM \cite{Jos}. 
Our final result reads
\be
\langle -| \, \Psi\rangle \sim (\partial^2)^4\, \Bigl [ \, 
(\partial\cdot\bar\partial)^2 - \partial^2\, \bar\partial^2\,
\Bigr ]^2 \, \frac{1}{(z\cdot \bar z)^8} 
= (\partial^2)^6\, (\bar\partial^2)^2\, \frac{1}{(z\cdot \bar z)^8} 
\, ,
\ee
which is {\it non-vanishing} and completes the proof of the non-triviality
of \eqn{ansatz}.

\vspace{6mm}
\noindent
{\bf Acknowledgement}

J.H. would like to thank M. Bordemann and R. Suter for previous collaborations
on the subject, J.P. thanks R. Helling and H. Nicolai for valuable discussions.

\end{document}